\documentclass[runningheads]{llncs}
\usepackage{graphicx}
\usepackage{amsmath,amssymb} 
\usepackage{color}
\usepackage[width=122mm,left=12mm,paperwidth=146mm,height=193mm,top=12mm,paperheight=217mm]{geometry}
\begin{document}

\pagestyle{headings}
\mainmatter
\def\ECCV18SubNumber{588}  

\title{Organ At Risk Segmentation with multiple modality} 


\author{
Kuan-Lun Tseng,
Winston Hsu,
Chunting Wu,
Ya-Fang Shih,
Fan-Yun Sun
}

\institute{National Taiwan University\\\texttt{r04921120@ntu.edu.tw}}

\maketitle

\begin{abstract}
With the development of image segmentation in computer vision, biomedical image
segmentation have achieved remarkable progress on brain tumor segmentation
and Organ At Risk (OAR) segmentation. However, most of the research only uses single modality 
such as Computed Tomography (CT) scans while in real world scenario doctors often use multiple modalities 
to get more accurate result.
To better leverage different modalities, we have collected a large dataset consists of 136 cases with
CT and MR images which diagnosed with nasopharyngeal cancer.
In this paper, we propose to use Generative Adversarial Network to perform CT to MR transformation
to synthesize MR images instead of aligning two modalities. The synthesized MR can be jointly trained with CT 
to achieve better performance.
In addition, we use instance segmentation model to extend the OAR segmentation task to segment both organs and tumor region.
The collected dataset will be made public soon.

\end{abstract}

\section{Introduction}
In medical image segmentation, many clinical workflows requires precise region of multiple organs or objects such as diagnostic interventions and radiation treatment planning.
In Radiation treatment planning, the delineation of Organs at Risk (OAR) in CT images is crucial for preventing irradiation of healthy organs.
Furthermore, an accurate segmentation of OAR can even support surgical planning such as image-guided neurosurgery.
However, manual segmentation of organs is time consuming and the results may vary from one doctor to another.
As a result, several automatic image segmentation algorithms are invented \cite{campadelli2009automatic} \cite{he2015fully}. 
Nowadays, most of the research focuses on CT images because of the clinical prevalence. However,
MR images are extremely informative toward tumor and soft tissue.
In practice, doctors usually fuse multiple modality such as MR, PET and CT to get more information
when some structures that have poorly visible boundaries in CT. Therefore, we want to incorporate
MR images to better segment OAR and tumor in head CT. 

\begin{figure}[t]
\begin{center}
\includegraphics[width=0.95\columnwidth]{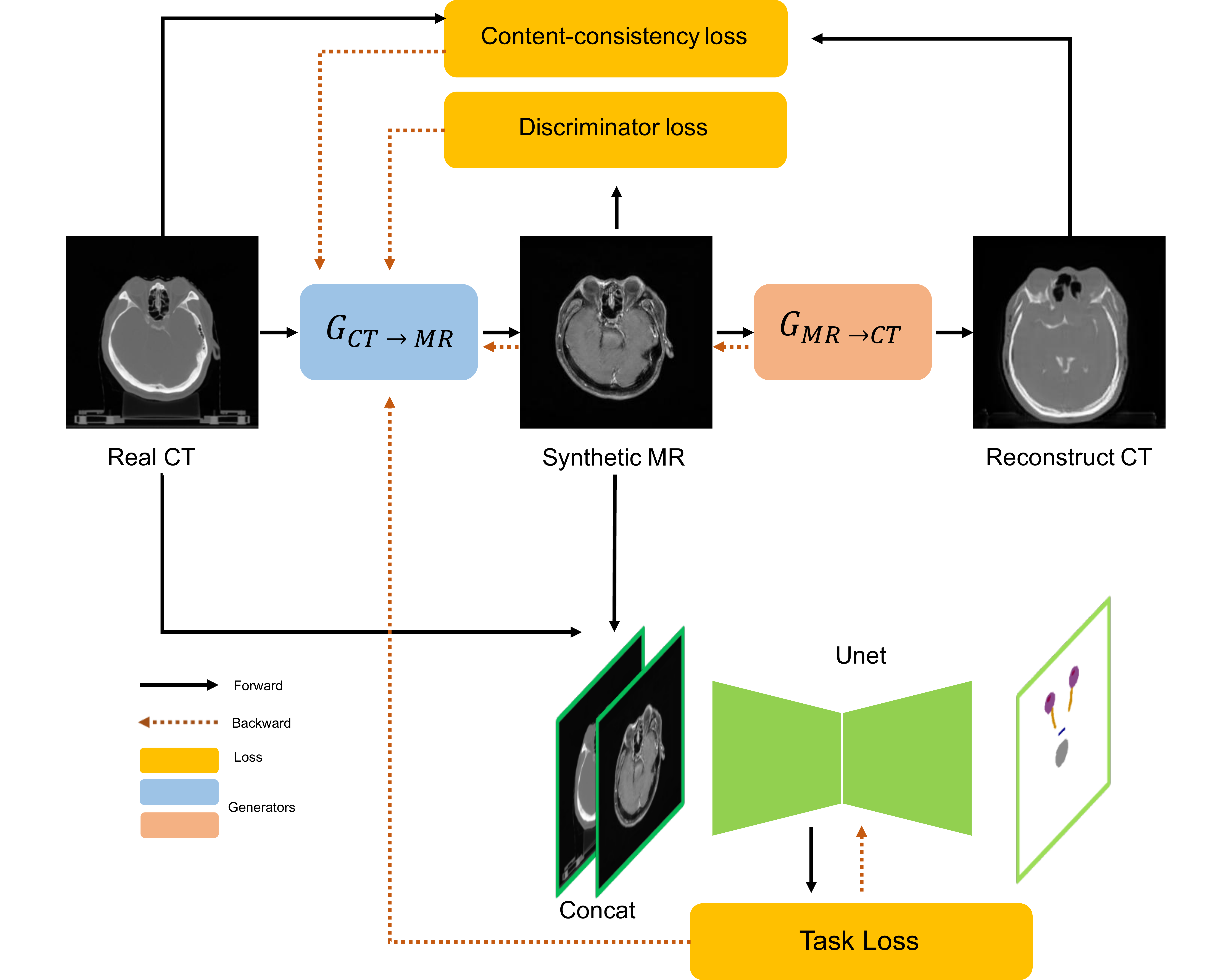}
\end{center}
\caption{
The illustration of our method from the CT to MR. Generators try to learn cross-domain translation
while regularized by content-consistency loss from original input and reconstructed source image and task loss from segmentation network.
Note that the backward cycle is omitted for clarity.
}
\label{figure:fig1}
\vspace{-0.12in}
\end{figure}

To better leverage CT and MR, we investigate the image fusion technique which proven to work very well in practice \cite{alexander1995magnetic}. 
Fusion of MR and CT can take the geometric precision of CT while eliminating the spatial distortion caused by susceptibility artifacts and peripheral magnetic field warping. 
However, most of the fusion algorithms need the modalities to be well aligned which could be very difficult in some cases.
We propose to use generative adversarial networks (GANs) to generate images from one modality to another while preserving its geometric precision (See Figure \ref{figure:fig1}).
With the recent advances in GANs, we can further alleviate the requirement for paired training which offers much more flexibility.

Our goal is to develop an automatic algorithm that can be integrated into current treatment planning process to
assist doctors. However, public biomedical image dataset is hard to find and many research [cite] use
their private dataset therefore is hard to compare and analyze.
As a result, we collaborate with Chung-Shan Medical University to collect 136
patients’ CT and MR scans (some fig). All the cases were used for Nasopharyngeal Carcinoma treatment planning 
and each CT scans have two sets which are taken before treatment and after treatment. 
Therefore we have total 272 scans can be used for training and testing. This dataset is quite large compared to existing
public available datasets. Every case have correspond MR image taken at the same time as the first CT.
The MR images are used to cross-reference the CT scans to better distinguish the tumor 
region and nerve cells. Note that the human label only exists in CT scans since CT scans are the primary resource for existing treatment planning system. 
We consulted with the exports and selected 10 major organs to predict, which are 
Brain-Stem, Chiasm, Cochlea, Eye, Inner Ears, Larynx, Lens, Optic Nerve, Spinal-Cord and GTV. 

Class imbalance is a major problem for biomedical image segmentation since most of the tissue are healthy or irrelevant.
This problem also exists in our dataset (Table \ref{table:stats}) thus directly training yield unsatisfactory results.
The major difficulty for semantic segmentation is that it lacks of instance information which cause
the boundary of nearby classes are difficult to separate. Another problem is that all classes
contribute equally to the segmentation loss therefore the loss is dominated by easy classes.
If we can incorporate spatial context each organ, segmentation across classes can avoid competing with each other.
Considering the advantage of instance segmentation and the fact that bounding box annotation can 
be obtained naturally when we already have pixel level annotations, we adopt Mask-RCNN \cite{he2017mask}
as our base model for this task. We investigate several design choices and exploit different loss functions to tackle class imbalance.

Incorporating with different modality data such as CT and MR is another challenge.
Aligning different modality data can be problematic if there is no additional knowledge such as human anatomy.
In our dataset, MR images were not aligned with any existing software and have different resolution, causing
direct use MR image as another modality to be impossible. A common strategy is to align each CT with MR using
mutual information. However, CT and MR images are sliced with different thickness therefore using
mutual information still cause sever mis-alignment. To solve the aligning difficulty, we opt to use
CycleGAN \cite{zhu2017unpaired} to synthesize MR image through CT.
With our modifications, such generative model can transform MR specific information to CT and preserve original
CT image structure. The aligned CT and Synthesized MR allows us to perform feature fusion with current state-of-the-art methods to get better performance.

\begin{table}[]
\centering
\resizebox{\textwidth}{!}{%
\begin{tabular}{l|c|c|c|c|c|c|c|c|c|c}
Category                     & BrainStem & Chiasm & Cochlea & Eye  & InnerEar & Larynx & Lens & OpticNerve & SpinalCord & GTV   \\ \hline
Number of images               & 3422      & 390    & 523     & 1642 & 376      & 1113    & 382  & 537        & 9693       & 4899       \\ \hline
Number of instances            & 3422      & 488    & 785     & 3062 & 551      & 1113   & 588  & 883       & 9693      & 5556 \\ \hline
\multicolumn{1}{c|}{median relative area} & 1.4 & 0.23 & 0.21 & 1.08 & 0.23 & 2.4 & 0.06 & 0.45 & 0.29 & 1.04 \\ \hline
\end{tabular}}
\bigskip
\caption{{\bf Statistic of our OAR dataset.} Relative area is computed as
the ratio of contour area over whole image area}
\label{table:stats}
\end{table}

\section{Related Work}

{\bf Biomedical Image Segmentation. }
There have been much research work that adopted deep methods for biomedical segmentation.
Havaei et al. [10] split 3D MRI data into 2D slices and crop small patches at 2D planes. 
They combine the results from different-sized patches and stack multiple modalities as different channels for the label prediction. 
V-Net \cite{milletari2016v} consists of a contracting path that contains multiple 3D convolutions for downsampling, and a expansive path that has several deconvolution layers to up-sample the features 
and concatenate the cropped feature maps from the contracting path. Fully utilize different resolution feature maps and
adopt dice loss to counter class imbalance. However, the proposed dice loss is not suitable for multi-class problems.
With the success of dice loss in binary segmentation task, Wang et al. \cite{wang2017automatic} proposed a cascade model structure
to deal with multi-label task. They combine tumor labels as one class while background as another,
and trained the model with binary segmentation fashion using dice loss. 
In second stage, they further split the different tumor classes into two classes and follow the same procedure as previous stage.
The network of each stage is trained separately and architecture are slightly different.
In this fashion, label imbalance problem will be further alleviated because the rare class has been located by other easy detected class.
Although the cascade structure indeed solved imbalance problem well, it is not practical since we have much more classes to segment.
\newline\newline{\bf OAR segmentation. }
Due to the success of deep learning based framework of natural images, segmentation of organs at risk often 
adopt CNN as their feature extractor. Recently Ibragimov et al. \cite{ibragimov2017segmentation} have proposed to extract
positive patches belong to the OAR of interest and negative patches around the surrounding structures.
They use those patches to train CNN for voxel classification and segment OARs in the test image where the corresponding OAR is expected to be located.
This pipeline involves patch training which is less efficient in both training and inference. 
Also, lots of domain knowledge was involved when selecting patches and OAR regions.
Several works \cite{trullo2017joint} \cite{trullo2017segmentation} used more advanced method \cite{pinheiro2016learning} to directly segment
organ instances in thoracic CT. In their experiment, the interested label are only four and thoracic CT has less diverse cell density
compare to Head CT. 
\newline\newline{\bf Instance segmentation. }
Instance segmentation methods are the joint task of detection and segmentation. 
DeepMask \cite{pinheiro2015learning} and SharpMask \cite{pinheiro2016learning} generate
segment proposals and then feed into RCNN to predict the class of the segmented region.
Li et al. \cite{li2016fully} introduce position-sensitive map to allow FCN network to perform detection.
This work can segment object even if the region proposal is fail.
Recently, He et al. \cite{he2017mask} proposed a simple framework which attach small mask prediction branch within the original 
Faster-RCNN network. Mask-RCNN decoupled the class prediction and mask prediction which prevent the class competition.
Therefore the class imbalance problem can be resolved as long as the region proposal is accurate.
\newline\newline{\bf Medical Image Synthesis. }
Image synthesis is an common approach in biomedical image analysis since real data and annotation is hard to collect.
With the development of GANs \cite{goodfellow2014generative} \cite{arjovsky2017wasserstein} \cite{berthelot2017began}
, many models have been successfully applied to medical images.
In \cite{nie2017medical} synthesizing MR images from CT using Generative Adversarial Networks because collecting MR 
is safer for patients. Cross-modalities synthesis can be used as domain adaptation problem \cite{kamnitsas2017unsupervised}.
While \cite{mahmood2017unsupervised} uses a reverse approach to translate real image to synthesized image
and feed to the networks trained on large datasets of synthetic medical images. 
However, the above approach are required for paired training data which is hard to collect especially with different modalities.
More recent research \cite{wolterink2017deep} \cite{chartsias2017adversarial} uses unpaired cross-domain trainable model CycleGAN \cite{zhu2017unpaired} to translate CT and MR images
. The cross-domain synthesized image are used as augmented training data to prevent the segmentation network from over-fitting.

\begin{figure}[t]
\begin{center}
\includegraphics[width=0.95\columnwidth]{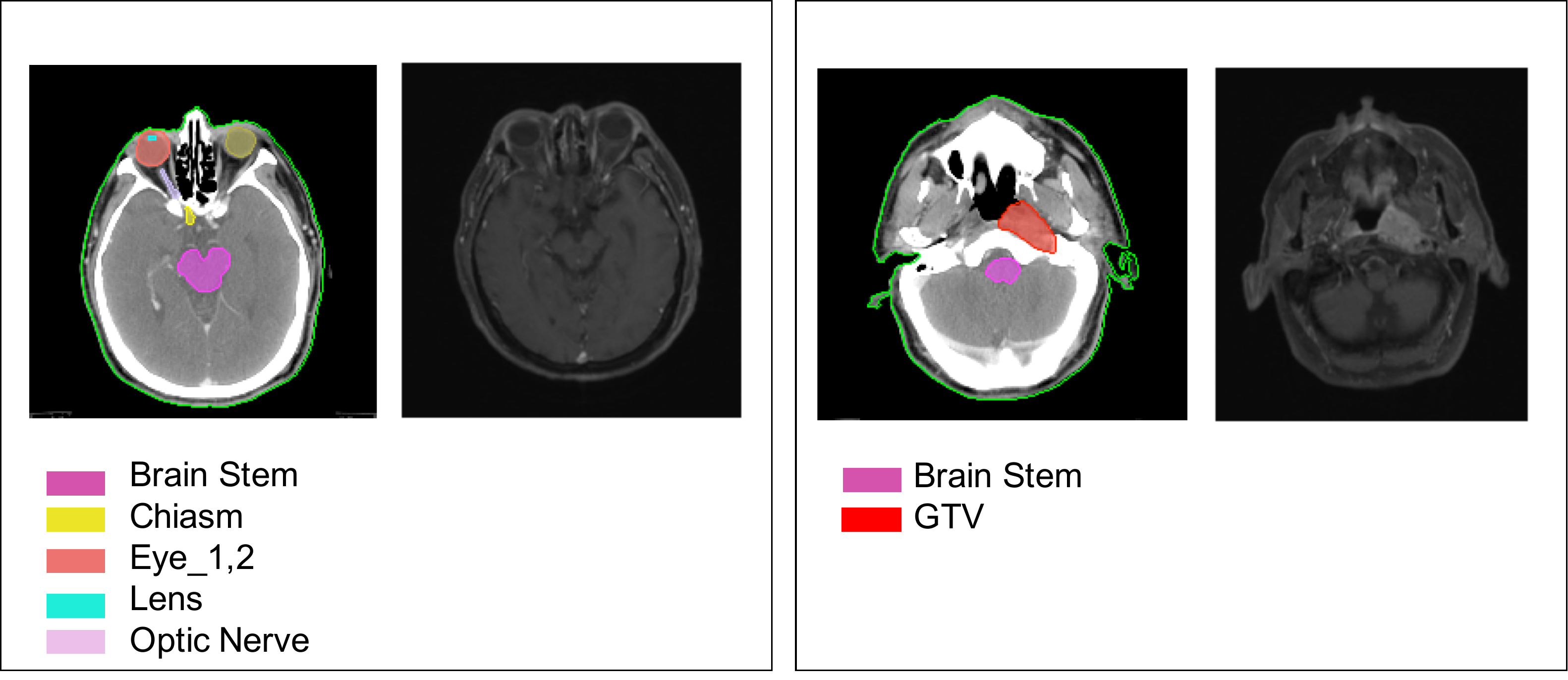}
\end{center}
\caption{
The left image of each block is the original CT scan and the organ contour
is highlighted by different color. Note that Eye and Optic Nerve were separately labeled by radiologist.
We integrated those labels as single class in all our experiments. The right image is the MR image of the same patient.
Noted that CT is not aligned with MR, the figure is manually selected for visualization purpose.
Best viewed in color.
}
\label{figure:fig2}
\vspace{-0.12in}
\end{figure}

\begin{figure}[t]
\begin{center}
\includegraphics[width=0.95\columnwidth]{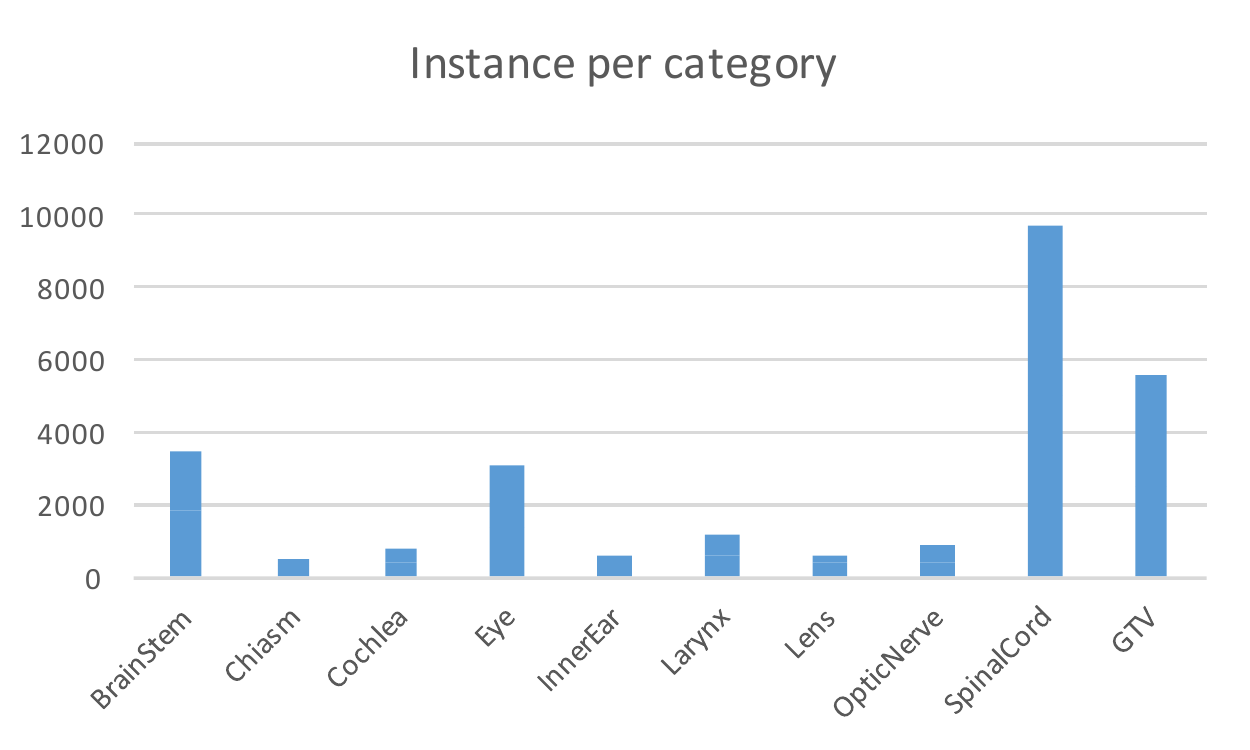}
\end{center}
\caption{Number of instances in each class.}
\label{figure:numInstance}
\vspace{-0.12in}
\end{figure}

\begin{table}[]
\centering
\begin{tabular}{|l|l|l|c|c|}
\hline
                    & cases & images                 & multi-modality & multi-label \\ \hline
Lung CT             & 60    & 9596                   &                & V           \\ \hline
Pancreas-CT         & 82    & 19,328                 &                & V           \\ \hline
Soft-tissue-Sarcoma & 51    & 38,283                 & V              &             \\ \hline
MM-WHS              & 60    & \multicolumn{1}{c|}{-} & V              & V           \\ \hline
Ours                & 272   & 29,422                  & V              & V           \\ \hline
\end{tabular}
\bigskip
\caption{Comparison between CT segmentation dataset}
\label{table:dataset}
\end{table}

\section{Dataset}
The CT images in the dataset were manually annotated by radiological experts in slice-by-slice fashion. 
Table \ref{table:stats} shows the statistics of the dataset. The number of instances of each organ class, the number of slices containing each class, and the median of relative area (the ratio of the bounding box area over the whole slice) of all instances in each class are reported. 
There are two features in our dataset:
first, the number of instances in each class vary largely. The instances and images of Spinal Cord is nearly 30 times larger than Chiasm (See Figure \ref{figure:numInstance} ).
Second, the organs are usually very small, only take up small area in each scan. Most of the median of relative area of all instances in every class are below one percent.
\subsection{Related Dataset}
There are some datasets that provide pixel-level annotations but usually contain only single modaity.
Lung CT Segmentation Challenge provides OAR segmentation dataset with only 60 sets of thoracic CT. 
While Cancer Imaging Archive Pancreas-CT dataset contains 82 abdominal contrast enhanced 3D CT scans from 53 male and 27 female subjects. Both of them have only single modality.
Soft-tissue-Sarcoma \cite{vallieres2015radiomics} provides 51 patients with histologically proven soft-tissue sarcomas (STSs) of the extremities.
All patients had pre-treatment FDG-PET/CT and MRI scans. Although there are multiple modality but only STS label are provided.
MICCAI 2017 Multi-Modality Whole Heart Segmentation (MM-WHS) dataset is most relevant to ours. The challenge provided 60 CT and 60 MRI volumes with corresponding
manual segmentation of seven whole heart substructures. Although they are multi-modality and multi-label, the size of the dataset is too small
for current deep learning methods. The detailed comparison is listed on Table \ref{table:dataset}
\subsection{Annotation}
The collected CT scans are provided with organ contours labeled by doctors for radiotherapy treatment planning.
Note that the annotation only exists in CT scans no annotation exists in MR images. The reason is that MR images are used as
cross-reference source and will not be used to calculate radiation dose. The categories of annotation in CT scans
are listed in Table \ref{table:stats}.
We extract contours from Dicom-RT file and calculate the minimum axis-aligned bounding box. After that we keep the center of the
bounding box and enlarge both width and height by a factor of 1.2 to avoid the overlap between the edge of the mask annotation 
and bounding boxes.Finally, We discard the annotation if the annotated area is fewer than 10 pixels.
After the first stage of label extraction, we need to verify the validity of the label. Labels in post-treatment CT 
are more noisy because tumor region may not visible after the first stage of treatment. As the result, radiologist may
only label important area. We refined 100 cases with experts and denoted as clean set. The remaining 86 cases preserve the original label
which some label may missing. We think that the noise in medical image annotation is inevitable since manually cleaning is time consuming.
Learning from noisy data is also an important research area we want to explore.

\subsection{Dataset Split}
We split our dataset into training and testing. All 30 testing cases are refined by experts. The remaining data
are used as training set. In the training set, only 70 cases are refined. Note that The pre-treatment and post-treatment scans of the same patient
are in the same set.
\subsection{Magnetic Resonance Imaging}
There are 186 sets of MR images associated with each patient taken before treatment. 
The thickness of slices ranges from 5.0 mm to 6.0 mm. Although the purpose of MR images is primary for diagnosing instead of localizing tumor region,
doctors usually fuse CT and MR if the detailed structure is not visible for CT.
\subsection{Computed Tomography Scans}
Each patient has two sets
of CT scans which were taken before treatment and after treatment. The post-treatment scans
are used for efficacy assessment and the radiation volume will be reduced after delivery of initial planned dose. 
As the results, the annotations in those scans is more likely to be noisy.

\begin{figure}[t]
\begin{center}
\includegraphics[width=0.95\columnwidth]{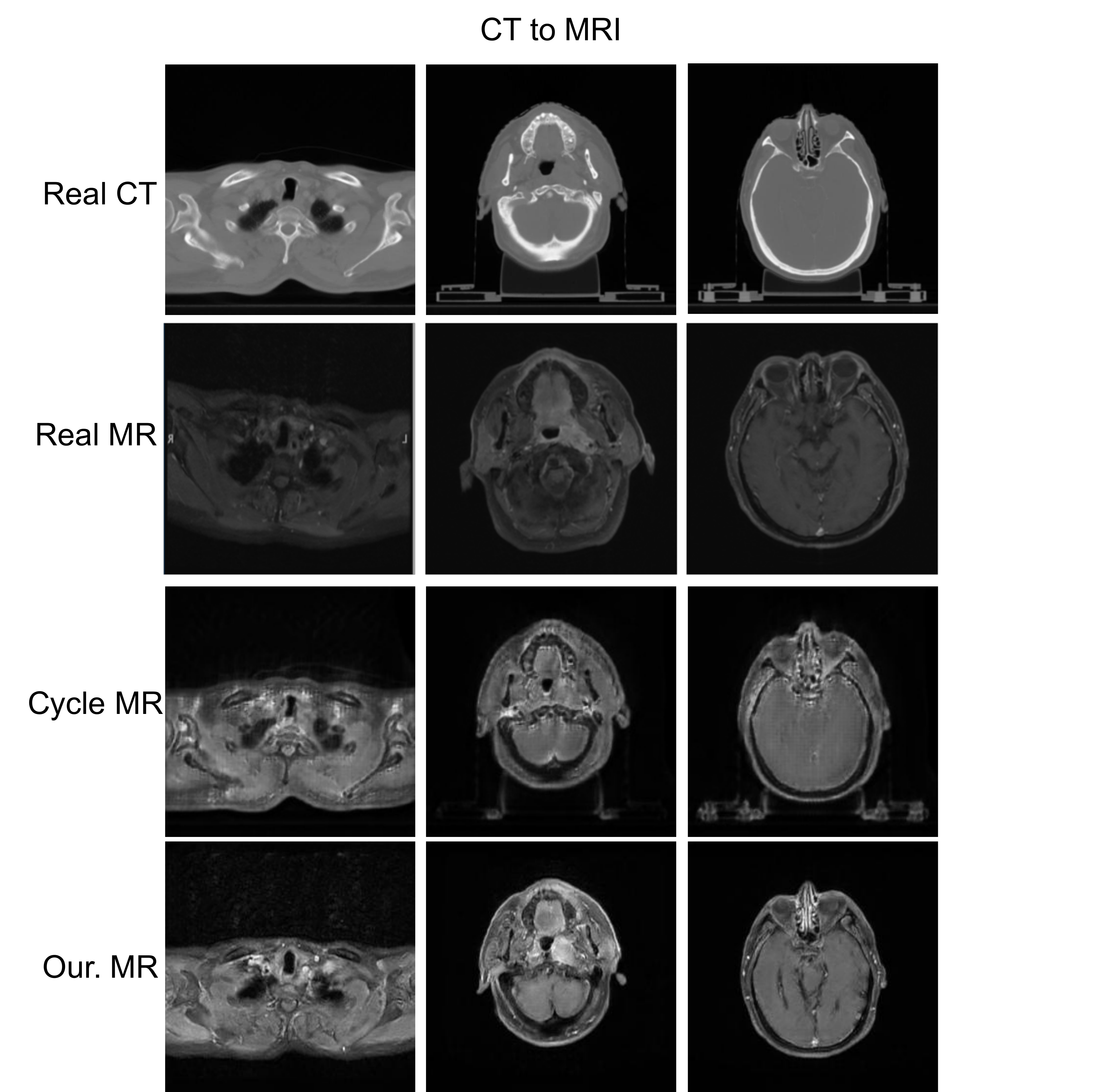}
\end{center}
\caption{
Example output for our MR generation results.
}
\label{figure:fig3}
\vspace{-0.12in}
\end{figure}

\section{Method}
Our OAR segmentation system contains two part: the  
MR image generation adopted CycleGAN \cite{zhu2017unpaired} with additional 
task loss and content similarity loss to generate MR image based on given CT.
The semantic meaning of the generated MR is well preserved and beneficial for learning.
The instance segmentation network take CT and generated MR as input to segment all interested organ.
\subsection{MR image generation} 
Since training with pairs of spatially aligned MR and CT is impossible in our dataset, we
adopt CycleGAN \cite{zhu2017unpaired} to translate image from one domain to another domain without unpaired training data. However, the default setting cannot produce satisfactory
result. As shown in Figure \ref{figure:fig3} column 3, CycleGAN synthesized MR is blurry and suffers from checkerboard artifacts.
\newline\newline{\bf Checkerboard Artifact.} To mitigate the checkerboard artifacts, we change resnet encoder into UNet since
skip connection can fuse semantic and structure information. Suggested by \cite{odena2016deconvolution}, we
change deconvolution layer in the UNet to nearest neighbor upsampling operation.
%
\newline\newline{\bf Content-consistency Loss.}
In Generative Adversarial Networks, the generator loss is defined as
\begin{multline} \label{eq:1}
\begin{split}
L_{GAN}(G_{S \rightarrow T}, D_{T}, X_{T}, X_{S}) = & \mathbb{E}_{x_{t} \sim X_{T}}[\log{D_{T}(x_{t})}]  \\ + &  \mathbb{E}_{x_{s} \sim X_{S}}[\log(1-D_{T}(G_{S \rightarrow T}(x_{s})))]
\end{split}
\end{multline}
In this objective function, the generator $G_{S \rightarrow T}$ will try to generate samples that looks like domain $T$.
While $D_{Y}$ tries to differentiate the generated sample $G_{S \rightarrow T}(x_{s})$ and real data $X_{T}$.
However, Equation \ref{eq:1} cannot ensure that the input $x_{s}$ will be mapped into desired output
with structure and content preserved. To reduce the space of possible mapping functions of two domain, cycle consistency
loss is introduced by \cite{zhu2017unpaired}. 
Since the goal of generated image is to assist source image segmentation, the region of interested labels are clearly more important than other tissue. 
We add another term that adds organ regions mask $M_{organ}$ to the L1 loss such that the organs are given more
weight than other tissue. Our content-consistency loss can be formualted as Equation \ref{eq:2}
\begin{multline} \label{eq:2}
L_{content}(G_{S \rightarrow T}, G_{T \rightarrow S}, X_{T}, X_{S}) = \\  \mathbb{E}_{x_{t} \sim X_{T}}[||(G_{T \rightarrow S} (G_{S \rightarrow T}(x_{s})) 
 - x_{s})\odot(1+M_{organ})||_{1}] \\ +   \mathbb{E}_{x_{s} \sim X_{S}}[||G_{S \rightarrow T} (G_{T \rightarrow S}(x_{t})) 
 - x_{t}||_{1}]
\end{multline}
Note that we can only apply content-consistency loss to regularize MR generator since we do not have MR annotation.
Content-consistency loss is analogous to content similarity loss \cite{bousmalis2017unsupervised} and pose mask loss \cite{ma2017pose}, which both aim to alleviate
the influence of background changes.
%
%
%
\newline\newline{\bf Task Loss.}
To further regularize the generators and guarantee the annotation correctness of source image, we
add another segmentation loss to encourage both generators to produce semantic-preserving samples. 
To this end, we concatenate the synthesized and real sample into an segmentation subnetwork $T$ and optimizing weighted cross-entrophy loss.
See Figure \ref{figure:fig1} green portion. The overall objective can be defined as:
\begin{multline} \label{eq:3}
\begin{split}
L(G_{S \rightarrow T}, G_{T \rightarrow S}, D_{T}, D_{S}, X_{T}, X_{S}, T) =  &L_{GAN}(G_{S \rightarrow T}, D_{T}, X_{T}, X_{S}) \\ + &L_{GAN}(G_{T \rightarrow S}, D_{S}, X_{S}, X_{T}) \\ + & L_{content}(G_{S \rightarrow T}, G_{T \rightarrow S}, X_{T}, X_{S}) \\ + &L_{task}(G_{S \rightarrow T},T)
\end{split}
\end{multline}
By alternatively optimizing the objective, generators will generate content consistant and semantic consistant images.
\subsection{OAR segmentation}
Mask RCNN is a multi-tasking model which simultaneously generate bounding boxes and object masks.
The network is based on Faster-RCNN \cite{ren2015faster} with additional mask branch to predict object mask based on bounding box.
This network is an perfect choice to our OAR segmentation because organs are rarely overlap with each other. The bounding box prediction
is farely simple since human organs' position is consistant. In addition, with the sampling stragey in Faster-RCNN, the label imbalancing problem which exists
in UNet will be alleviated. There are three subnetwork attached to each roi, which perform bouning box regression, classification and segmentation.
The segmentation loss is defined according to the roi associated ground-truth class. In this manner, Mask RCNN can generate masks for every class without competition among classes.

\section{Implementation Details}
We modified the Mask-RCNN and CycleGAN network architecture to better fit our task.
The training procedure and implementation are also different from original paper.
All the models are implemented using Tensorflow \cite{tensorflow2015-whitepaper}.
\subsection{CycleGAN Details}
{\bf Architecture.} Directly training CycleGAN with default encoder cannot produce high quality images
because of the checkerboard effect. Therefore we opt to use UNet with three convolution layers with stride two and add
skip connections at each symmetric level. We found that skip connection can lead to more stable training.
LeakyReLU and instance normalization are used in all convolutional layers for both generator and discriminator.
The initial featuremap channels of generator are 64 and multiply by 2 at each downsampling layer. 
The segmentation subnetwork described above is an UNet with 3 downsampling layers for memory efficiency.
We trained the segmentation subnetwork with weighted cross entrophy loss as in \cite{trullo2017joint} to alleviate the imbalance among classes.
Otherwise, the network will heavily biased toward background.
We calculated median-frequency weights \cite{eigen2015predicting} through the whole training data.
The median-frequency is defined as:
\[{\alpha _c} = median\_freq/freq(c)\]
where $freq(c)$ is the number of pixels of class $c$ divided by the total number of pixels in
images where $c$ is present, and $median\_freq$ is the median of all class frequencies.
Therefore the dominant labels will be assigned with the lowest weight which balances the training process.
\newline\newline{\bf Training Details.} We use Adam optimizer with learning rate 0.0002 and batch size 1 for all the training of CycleGAN and segmentation subnetwork.
The CT image is center cropped with $250 \times 250$ to remove backgoround, then CT nad MR images are first resized to $256 \times 256$ and fed into CycleGAN.
During training, we perform data augmentation with center crop and random flip.
\subsection{Mask RCNN Network Architecture}
{\bf Architecture.} In terms of backbone selection, we have experimented with resnet-50 and resnet-101 and found that no significant difference in terms of performane.
Therefore we choose resnet-50 to reduce memory usage and training time. we conduct experiment using resnet-50-C4 as in \cite{he2016deep} and FPN \cite{lin2017feature}. 
For FPN, we build top down and bottom up feature prymaid with lateral connection to the corresponding level. The feature map used by rpn start from stride 4 and stride 64 is discarded because the label in our
dataset is relatively small compared to natural image dataset. The mask branch resolution is $28 \times 28$ as the original paper since larger resolution yield
no improvement in our experiments.
\newline\newline{\bf Training Details.}
We use gradient decent optimizer with momentum 0.9 for 18 epochs. 
The initial learning rate is 0.001 and increase linearly to 0.01 during first 3 epochs.
After that, we decrease the learning rate by factor of 10 at 5 and 10 epochs.
This learning rate configurations are used in all experiments.
The positive and negative ratio are 1:3 and foreground thresholds are both 0.5 for RPN and RCNN.
Number of rois in each mini-batch is 256 and each mini-batch has only single image per GPU.
All the Mask-RCNN experiments are trained with 2 NVIDIA P100 GPUs.
In training, we conduct more data augmentation because we observed sever over-fitting if only random flip is used.
Through experiments, we found that scale jittering is the most effective way. 
Input images are resized such that the size of shorter edge is randomly picked from {800, 900, 1100, 1200} pixels.

\begin{table}[]
\centering
\resizebox{\textwidth}{!}{%
\begin{tabular}{c|ccccccccc}
           & BrainStem & Chiasm & Cochlea & Eye  & InnerEar & Larynx & Lens & OpticNerve & SpinalCord \\ \hline
V-Net      & 0.61      & 0.0    & 0.07    & 0.58 & 0.02     & 0.37   & 0.06 & 0.01       & 0.66       \\
V-Net-WL   & 0.77      & 0.12   & 0.36    & 0.75 & 0.16     & 0.46   & 0.12 & 0.21       & 0.77       \\
V-Net-GDL   & 0.75      & 0.1    & 0.41    & 0.77 & 0.27     & 0.43   & 0.17 & 0.23       & 0.76       \\ \hline
Dense-V-WL    & 0.64      & 0.13   & 0.19    & 0.62 & 0.19     & 0.44   & 0.39 & 0.19       & 0.55       \\
Dense-V-FL & 0.79      & 0.12   & 0.42    & 0.76 & 0.33     & 0.49   & 0.18 & 0.32       & 0.77       \\ \hline
MaskRCNN   & 0.84      & 0.30   & 0.508    & 0.85 & 0.43     & 0.49   & 0.52 & 0.36       & 0.72      
\end{tabular}}
\bigskip
\caption{
{\bf Organ At Risk dice score on clean set.} Both V-Net \cite{milletari2016v} and Dense-VNet suffer from
label imbalance as we can find that the scores of Chiasm, Cochlea and InnerEar are low. We further apply
generalised dice loss (GDL), weighted cross-entrophy loss (WL) and focal loss (FL) and observe significant improvement. Instance segmentation
method MaskRCNN achieves best score as we expected. 
}
\label{table:baseline}
\end{table}

\section{Experiment}
We evaluate several state-of-the-art biomedical image segmentation models as our baseline.
The MR generation results of differnet network architecture are also discussed in the section.
In addition, we further analyze the class imbalance problem by using differnet loss such as 
focal loss \cite{lin2017focal} and generalised dice loss \cite{sudre2017generalised}. 
Finally we show the performance of using synthesized multi-modal data can actually improve performance.
\subsection{Evaluation Setup}
We use dice score obtained by each interested organ to evaluate our OAR segmentation performance. Dice score is a common
evaluation criteria for biomedical image segmentation. The Dice score is not only a measure of how many positives but it also penalizes for the false positives.
\[
Dice = \frac{P_1 \bigcap T_1}{(P_1 + T_1)/2}
\]
where T is ground truth label and P is predicted result.
$T_1$ is the organ area and $P_1$ is the pixels predicted as positives for the organ region.
\subsection{Representative Approaches}
{\bf V-Net. } 
We use the public available V-Net \cite{milletari2016v} implementation in NiftyNet \cite{niftynet17} as the comparing baseline model. 
The network use 3D convolutions to encode the correlation between slides and the structure is similar to U-Net. 
A typical V-net architecture consists of several convolution downsampling layers followed by transpose convolution upsampling stage. 
Skip connections are used to propagate the infomation from downsampling to upsampling stage to improve the final segmentation results.
\newline\newline{\bf Dense-Vnet.} A densely connected version of V-Net proposed in \cite{gibson2018automatic} is also trained as our second baseline. 
In this network, the input of each of the downsampling stage is the stack of feature from all the preceeding convolution block, which call a dense feature stack. 
The downsampling stage consists three dense feature stack follow by upsampling using bilinear upsampling layers to the final segmentation results. 
With this structure, we are able to extract more compact information from the 3D images and improve the segmentation results.
We trained networks on NVIDIA K80 GPU with batch size of 3 and the The input 3D patches is sampled from the CT images with spatial window size (64, 128, 128). 
Both of the networks are pre-trained using weighted softmax loss for 20k iterations, then continued the training using focal loss \cite{lin2017focal} for 20k iterations in order to address the class imbalance problem in our dataset.
The resulting segmentation masks are quantitatively evaluated in terms of Dice score and shown in Tabel \ref{table:baseline}.

\begin{table}[]
\centering
\resizebox{\textwidth}{!}{%
\begin{tabular}{c|ccccccccc}
             & BrainStem     & Chiasm        & Cochlea       & Eye           & InnerEar      & Larynx        & Lens          & OpticNerve    & SpinalCord    \\ \hline
MaskRCNN-CT  & 0.84          & 0.30          & 0.50          & 0.82          & 0.43          & 0.49          & 0.52          & 0.36          & 0.72          \\ \hline
Fusion@I     & \textbf{0.86} & 0.30          & 0.54          & \textbf{0.87} & 0.45          & \textbf{0.61} & \textbf{0.55} & 0.41          & \textbf{0.74} \\
Fusion@F     & \textbf{0.86} & \textbf{0.32} & 0.56          & 0.85          & 0.58          & 0.56          & 0.53          & \textbf{0.49} & 0.73          \\
Fusion@I-FPN & 0.85          & 0.31          & \textbf{0.60} & 0.85          & \textbf{0.59} & \textbf{0.61} & 0.53          & 0.48          & 0.73         
\end{tabular}}
\bigskip
\caption{
{\bf MaskRCNN results with different fusion scheme.} MaskRCNN-CT means that using only CT as input. Fusion@I and Fusion@F take both
CT and synthesized MR as input, the former concatenates two images alone the last channel and the latter split into two branch (form conv1 to conv4)
and then concatenate together. Two types of fusion scheme yield similar results. Noted that Fusion@F is hard to integrate into FPN archetecture because
of the memory issue and since the performance is similar, we opt to use Fusion@I for FPN backbone.
}
\label{table:fusion}
\end{table}

\subsection{Main Results}
{\bf Class Imbalance.}
In our experiments, VNet and Dense-VNet are hard to train with softmax cross entrophy. Both of them fail 
to segment rare classes such as Chiasm and Cochlea (See Tabel \ref{table:baseline}). Next we exploit weighted-cross entrophy loss which the balancing weights
are computed according to median frequencies. Both dominant classes and rare classes are significantly improved. We further investigate
generalised dice score to directly optimize our criteria and found that overall classes are improved.
To better learning the hard classes we opt to use recent proposed focal loss which encourages network put more attention to learn hard classes. However, the training is not stable at the begining
and quickly diverged, thus we trained focal loss from weighted softmax pre-trained models and observe large performace gain.
Despite the effort we put into solving class imbalance, including two stage training and manually selecting weights, it is still hard to get improvement on all classes.
The MaskRCNN uses different stragey which segmentation is only performed on positive rois and uses sigmoid to avoid class competition. This pipeline can
handle the imbalance of background class gracefully since positive roi must contain interested objects with certain IOU. Experiments show that MaskRCNN achieves best scores on all classes without bells and whistles (See Table \ref{table:baseline}).
\newline\newline{\bf Multi-Modality training.} 
Incorperating multiple modaities into training has proven effective in biomedical images \cite{alexander1995magnetic} \cite{guo2017medical}.
We experiment two fusion scheme: fusion at input and fusion at featuremap. We denote concatenating CT and synthetic MR along channel dimension as Fusion@I.
Using two branches start from conv1 to conv4 is denoted as Fusion@F. Finally we report scores using FPN backbone.
As we can see from Table \ref{table:fusion}, all the fusing scheme yield significant improvement.
\begin{table}[]
\centering
\resizebox{\textwidth}{!}{%
\begin{tabular}{c|cccccccccc}
                                & BrainStem     & Chiasm        & Cochlea       & Eye           & InnerEar      & Larynx        & Lens          & OpticNerve    & SpinalCord & GTV           \\ \hline
\multicolumn{1}{l|}{Dense-V-DL} &               &               &               &               &               &               &               &               &            & 0.39          \\ \hline
MaskRCNN-CT                     & \textbf{0.87} & 0.22          & 0.45          & 0.86          & 0.34          & 0.60          & 0.57          & 0.44          & 0.75       & 0.55 \\
Fusion@I-FPN                    & 0.86          & \textbf{0.29} & \textbf{0.56} & \textbf{0.87} & \textbf{0.55} & \textbf{0.61} & \textbf{0.58} & \textbf{0.49} & 0.75       & \textbf{0.56}      
\end{tabular}}
\bigskip
\caption{
{\bf GTV segmentation results.} Dense-VNet with binary dice loss is reported at first column.
As we can see, MaskRCNN wins by a large margin and achieves the same performance as trained only on OAR.
}
\label{Table:GTV}
\end{table}
\newline{\bf Simultaneously segment OAR and Tumor.}
We further extend OAR segmentation to simultaneously segment GTV.
Different from perform OAR segmentation, GTV region may overlap with OAR region and both information are need in treatment planning.
Since semantic segmentation models can only ouput single class for each pixel, the overlapped regions will be ignored.
In the experiment, we train Dense-VNet solely on GTV label to avoid the aforementioned issue and shows that instance segmentation method can do 
OAR and GTV at the same time without performance drop (See Table \ref{Table:GTV}).
\section{Conclusions}
In this work, we have introduced an new dataset for Organ At Risk segmentation.
We have demonstrated the key features of our dataset: an challenging multiple modalities and sever label imbalance.
We have conducted in-depth analysis on current state-of-the-art biomedical segmentation methods and proposed a effective way to use
synthesized MR while preserving semantic meaning. In addition, we have showed that joint learning with multiple modaities can significantly improve segmentation results. 

\bibliographystyle{splncs}
\bibliography{eccv2018submission}
\end{document}